\documentclass[times,authoryear]{elsarticle}
\usepackage{jasr}
\usepackage{myaasmacros}
\usepackage{framed,multirow}

\usepackage{amssymb}
\usepackage{amsmath}
\usepackage{latexsym}
\usepackage{graphicx}	% Including figure files
\usepackage{caption}
\usepackage{subcaption}
\usepackage{gensymb}
\usepackage{multirow}
\usepackage{xspace}
\usepackage{siunitx}
\usepackage[normalem]{ulem}

\usepackage[switch]{lineno}

\usepackage{url}
\usepackage{xcolor}
\definecolor{newcolor}{rgb}{.8,.349,.1}

\usepackage[citebordercolor=white]{hyperref}

\journal{Advances in Space Research}

\newcommand{\hanna}[1]{{#1}}

\begin{document}

\verso{A.V. Popkova \textit{etal}}

\begin{frontmatter}

\title{Search for H$_2$ cloudlets in our backyard}%Type the title of your paper, only capitalize first
%word and proper nouns\tnoteref{tnote1}}%
%\tnotetext[tnote1]{This is an example for title footnote coding.}

% [Popkova et al.]{
% A.V. Popkova, $^{1,2,5}$\thanks{E-mail: av.fokanova@physics.msu.ru}
% M. S. Pshirkov,$^{1,2,3,5}$
% A.V. Tuntsov$^{4}$
\author[1,2,4]{A.V. \snm{Popkova}}%\corref{cor1,cor2,cor5}}
\ead{av.fokanova@physics.msu.ru}
% \cortext[cor1]{Corresponding author: 
%   Tel.: +0-000-000-0000;  
%   fax: +0-000-000-0000;}
\author[1,2,3,4]{M.S. \snm{Pshirkov}}
% \fntext[fn1]{This is author footnote for second author.}
\author[5]{A.V. \snm{Tuntsov}}
%\address[2]{Affiliation 1, Address, City and Postal Code, Country}

\affiliation[1]{organization={Sternberg Astronomical Institute, Moscow  M.V. Lomonosov State University},
                addressline={Universitetskiy Prospekt, 13},
                city={Moscow},
                postcode={119992},
                country={Russia}}

%\address[2]{Affiliation 2, Address, City and Postal Code, Country}
\affiliation[2]{organization={Faculty of Physics, Moscow M.V. Lomonosov State University},
                addressline={Leninskie gory 1},
                city={Moscow},
                postcode={119991},
                country={Russia}}
                
%\address[3]{Affiliation 3, Address, City and Postal Code, Country}
\affiliation[3]{organization={Lebedev Physical Institute, Astro Space Center, Pushchino Radio Astronomy Observatory},
                addressline={Radioteleskopnaya 1a},
                city={Pushchino, Moscow reg.},
                postcode={142290},
                country={Russia}}

%\address[4]{Affiliation 4, Address, City and Postal Code, Country}

%\address[5]{Affiliation 5, Address, City and Postal Code, Country}
\affiliation[4]{organization={The Institute for Nuclear Research of the Russian Academy of Sciences},
                addressline={B-312, Prospekt 60-letiya Oktyabrya, 7a},
                city={Moscow},
                postcode={117312},
                country={Russia}}    
\affiliation[5]{organization={Manly Astrophysics},
                addressline={15/41-42 East Esplanade},
                city={Manly},
                postcode={NSW 2095},
                country={Australia}}                 
\received{July 2024}
\finalform{10 May 2013}
\accepted{13 May 2013}
\availableonline{15 May 2013}
\communicated{S. Sarkar}

\begin{abstract}
%%%
Several observational lines of evidence imply that a fraction of the dark matter in the Galaxy may be comprised of small cold clouds of molecular hydrogen. Such objects are difficult to detect because of their small size and low temperature, but they can reveal themselves with gamma radiation arising in interactions between such clouds and cosmic rays or as dark shadows cast on the optical, UV and X-ray sky background.  In our work we use the data of Fermi LAT 4FGL-DR4 catalogue of gamma-ray sources together with the data of GALEX UV All-Sky Survey to search for small dark clouds of molecular hydrogen in the Solar neighbourhood. This approach allows us to put an upper limit on the local concentration of such objects:  $n < 2.2 \times 10^{-11} {\mathrm{AU}^{-3}}$. Constraints (upper limits) on the total amount of matter in this form bound to the Sun strongly depend on the radial profile of the clouds' distribution and reside in \hanna{$0.05-30~M_{\odot}$} mass range.
%%%%
\end{abstract}


\end{frontmatter}

% %% For linenumbers
 %\linenumbers

\section{Introduction}
Despite sustained effort, our understanding of the elusive dark matter (DM) in the Universe and the Galaxy is far from being complete. One of the main issues is  the composition of the DM: although different lines of evidence imply that the majority of it consists of some non-baryonic substance \citep{nonbaryonic2, nonbaryonic3, nonbaryonic4}, there could be considerable contribution from ordinary matter as well \citep{barDM4,barDM3,barDM1}.
One particular suggestion has been self-gravitating clouds of cold, dense gas, predominantly molecular hydrogen. They are an attractive baryonic DM candidate as they can naturally resolve the galactic cusp problem and efficiently tie star formation to the total dynamical mass \citep{1999MNRAS.308..551W}. A variety of observational phenomena can be explained by recourse  to existence of  such objects, including extreme scattering events \citep{2016Sci...351..354B}, intra-day variability of pulsars and compact extra-galactic radio sources \citep{2019MNRAS.487.4372B} or optical variability of field quasars \citep{1993Natur.366..242H, 2022MNRAS.513.2491T}. These observations can be best explained using the characteristic size of the cloud $\mathcal{O}(1)~$AU and characteristic mass around $10^{-5}~M_{\odot}$. Such dense clouds would be  difficult to detect because of their small size and low temperature ($\sim 10$ K). Molecular hydrogen remains dark at such temperature as  it's lines start to get excited only at the temperature of a few hundred K. 
 
 Cloud population could consist of unbound clouds and those gravitationally bound to stars, including the Sun \citep{2017ApJ...843...15W}. The number of tiny clouds bound to each star is estimated to be of the order of  $10^5$, which would make a dynamically important contribution to the local density, similar to the stellar component \citep{2017ApJ...843...15W}. 

Closest objects in a population of such clouds bound to the Sun could be prominent sources of $\gamma$-rays in the high energy range (>100 MeV), generated during collisions of the galactic cosmic rays with the hydrogen nuclei, so they could be detected as  some unidentified sources in the Fermi LAT \citep{Atwood} source catalog.
Another observational signature of such dense objects would be produced due to their opacity for photons of lower energies from UV to X-ray: a  cloud would cast  a potentially observable shadow on the background.
To search for this effect, we used UV-observations of the GALEX telescope \citep{galex}, given its almost full coverage of the sky and high angular resolution. Although the shadow size would be larger at X-ray energies, the angular resolution of all-sky mission such as ROSAT and Spektr-RG is considerably lower.

We searched  for distinctive shadows near positions of unidentified sources in the 4FGL-DR4 catalogue (see \cite{4FGLDR3,4FGLDR4}). Our methods allowed us to place constraints on the concentration of such clouds in the Solar neighbourhood, $n < 2.2 \times 10^{-11}$AU$^{-3}$. 

The paper is organized as follows. In Section~\ref{sec:model} we describe the model of a small molecular cloud that we used in our paper. In Section ~\ref{sec:data} we describe the data we analyzed and  present the limits obtained from them, in Section ~\ref{sec:discussion} we discuss these limits and possible caveats of our approach. We present the main results and conclude in Section ~\ref{sec:results}.

%%%%%%%%%%%%%%%%%%%%%%%%%%%%%%%%%%%%%%
%%%%%%%%%%% CLOUD MODEL %%%%%%%%%%%%%
%%%%%%%%%%%%%%%%%%%%%%%%%%%%%%%%%%%%%%
\section{Model of a molecular cloud}\label{sec:model}
\begin{figure}
\centering
\includegraphics[width=0.5\textwidth]{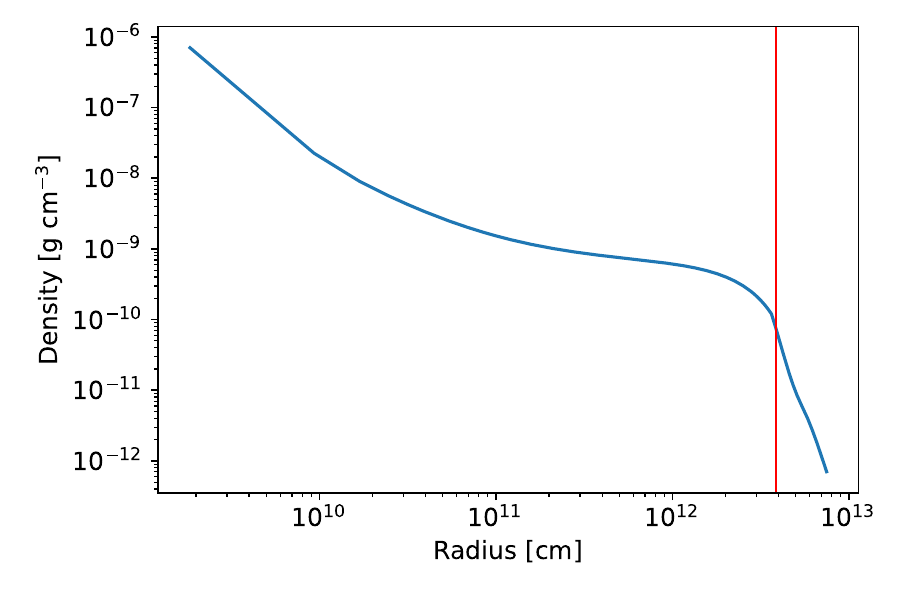}
\caption{Density profile of the model cloud. The vertical line indicates the approximate radius of the radiating shell (we assume that no cosmic ray particles would penetrate deeper, with all of them interacting with the nuclei inside the shell), which corresponds to the column density of hydrogen $\approx 40~ \mathrm{g/cm}^2$ }.
\label{fig:density}
\end{figure}

 Cold clouds can have a variety of possible internal structures depending on the composition, equations of state of the components, details of the physical processes maintaining the equilibrium as well as any assumed boundary conditions \citep{2019ApJ...881...69W}. The present paper is based on a model cloud in marginal convective equilibrium everywhere outside the metal core containing about one per cent of the total cloud mass. It uses realistic equations of state for both hydrogen \citep{2009JPCRD..38..721L} and helium \citep{2021JPCRD..50d3102P} in a 3:1 mass ratio but ignores non-ideal contributions to the entropy of mixing of the two fluids. This results in a characteristic mass and radius of $3\times10^{-5}~M_{\odot}$ and $\sim 0.5\,\mathrm{AU}$, placing it roughly in the middle of the range preferred by the quasar microlensing analysis of \cite{2022MNRAS.513.2491T}. The density profile of the cloud is presented in Fig. \ref{fig:density}.

\begin{figure}%[h!]
\centering
\includegraphics[width=0.5\textwidth]{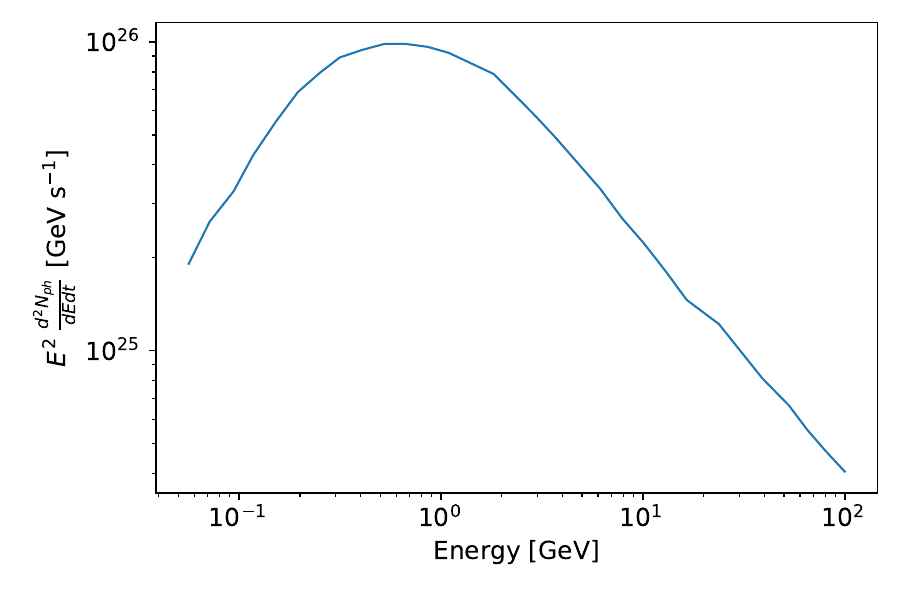}
\caption{Expected gamma-ray spectrum of the cloud}
\label{fig:spectrum}
\end{figure}

Interaction of the galactic cosmic rays (CRs)  with the clouds would lead to abundant production of $\gamma$-rays, mostly  due to proton-proton and proton-helium collisions. The expected spectrum of the signal would be very close to the spectrum of the nearby molecular clouds in Chameleon, Galactic anticenter and Orion-Eridanus superbubble \citep{2021Univ....7..141T} and would have a wide maximum at GeV energies (see Fig. \ref{fig:spectrum}).

The expected gamma-ray luminosity could be estimated using the known gamma-ray emissivity of hydrogen atom\footnote{For the sake of simplicity we ignore the difference between p-p and p-nucleon inside He  collisions and consider that the cloud consists of pure hydrogen.} $q(E)$ and the amount of the cloud material participating in the interactions. Gamma-ray emissivity per hydrogen atom is the amount of gamma-ray energy emitted by one hydrogen atom in gamma-ray photons of a particular energy E per unit time per unit solid angle. The dependence $q(E)$ of gamma-ray emissivity per hydrogen atom on the energy of the emitted gamma-ray photon in the local interstellar medium was taken from Fig.2 in \cite{emissivity}. Due to the high density of the cloud, most of its mass would be shielded from the CRs and we are  only interested in  the atoms residing in the shell with a width corresponding to the  CR interaction length in the hydrogen. The interaction grammage  is $40~\mathrm{g~cm}^{-2}$, assuming interaction cross-section $\sigma_{pp}=40$~mb, so with the adopted model the shell width  would be $ h \approx 0.2$ AU and we are essentially dealing  with a model of 'thick atmosphere around opaque planet'. The vertical line on the density profile of the cloud (see Fig. \ref{fig:density}) indicates the inner border of the shell. We adopt a simplified model that only matter  in this shell gives  contribution to the observed flux. An actual contribution from the deeper layers is suppressed by  two processes: first, due to geometrical reasons, most of the CRs would traverse grammage considerably larger than the shell width; second, the photon attenuation length at these energies is quite short, corresponding to grammage $\sim80~\mathrm{g~cm}^{-2}$, and that would lead to further suppression of flux from these layers. The shell mass would be $\int_{R-h}^{R} 4 \pi r^2 \rho(r) dr \approx 5.3\times10^{-6}~M{\odot}$, which corresponds to the number of hydrogen atoms $N_H \approx 6.4\times10^{51}$. We make another simplifying assumption that around the half of gamma-ray emission in our direction would get out of the shell, and another half coming from the far side would be absorbed in the inner regions. \hanna{The expected spectrum from the cloud  is presented in Fig. \ref{fig:spectrum}. The resulting luminosity  at the energies higher than 1 GeV that would be used  further can be estimated by numerical integration of this spectrum in the required energy range:}
\begin{equation}
L(E>1~\mathrm{GeV})\approx \frac{1}{2}\int_{1~\mathrm{GeV}}^{100 ~\mathrm{GeV}} 4 \pi q(E)    N_H dE \approx 6.8\times10^{25}~\mathrm{phot~s^{-1}}
    \label{eq:lum}
\end{equation}

%%%%%%%%%%%%%%%%%%%%%%%%%%%%%%%%%%%%%%
%%%%%%%%%%% GAMMA_RAY DATA AND CONSEQUENT LIMITATION %%%%%%%%%%%%%
%%%%%%%%%%%%%%%%%%%%%%%%%%%%%%%%%%%%%%
%\section{Gamma ray data and soft limitation on the concentration of molecular clouds}\label{sec:mild_limit}
\section{Data and limits}\label{sec:data}

Potentially this signal could be detected by the Fermi LAT \citep{Atwood} instrument. In this  case the cloud would be included in the Fermi LAT catalogue as an unidentified source. For our analysis we used the 4FGL-DR4 version of the catalogue \citep{4FGLDR4}, based on 14 years of the observations  in the energy range from 50 MeV to 1 TeV. The catalogue contains 7194  sources.

We made a sample of possible dark molecular cloud candidates from 4FGL-DR4 catalogue using the following criteria. We  selected only unassociated sources, because our clouds are  dark at other energies. Unassociated sources at low galactic latitudes are designated as 'unk' ('unknown'), while unassociated sources at high galactic latitudes  are designated as 'bcu' ('blazar galaxies of uncertain type') in 4FGL-DR4 catalog, i.e. all unassociated sources at high latitudes are bundled as blazars \citep{bcu}.  That is why we included both 'bcu' and 'unk' type objects into our sample--1772 unassociated sources in total. Then we selected the unassociated sources with high enough detection significance $> 5 \sigma $ and  also filtered the sources by the variability index presented in the catalog. The variability index is the sum of the log(likelihood) difference between the flux fitted in each time interval and the average flux over the full catalog interval; a value greater than 18.48 over 12 intervals indicates <1\% chance of being a steady source  \citep{var_index}. We assume that the CR flux is non-variable, neglecting solar modulation, as the clouds under consideration are located outside the heliosphere. As a result we expect the gamma-ray flux from dark molecular clouds to be non-variable as well, so we took only the sources with variability index less than 18.48. Finally, we made a spectral cut to eliminate the sources with the spectrum far from the expected one. The average cosmic-ray spectrum in the local Galaxy in the 1–100 GeV range is well described by a broken power-law in rigidity with a low-energy slope of $2.33^{+0.06}_{-0.08}$
 and a break at $18^{+7}_{-4}$ GV, with a slope change by $0.59 \pm 0.11$ \citep{spindex}. Taking this into account, we considered the sources with the spectral index from 2.0 up to 2.8 which is not too dissimilar to the  spectrum, expected from the clouds (see Fig. \ref{fig:spectrum}) .  Our final sample included 515 sources. 

The upper limit on the concentration of dark molecular clouds near the Sun can be obtained assuming that all the unassociated sources from the sample are dark clouds. The maximum distance of detection could be estimated from Eq. \ref{eq:lum} and the Fermi LAT sensitivity curves. 

We took the Fermi LAT sensitivity curves from LAT performance page (see
https:\slash \slash www.slac.stanford.edu\slash exp\slash glast\slash groups\slash canda\slash lat \textunderscore Performance.htm). In the energy range from 1 to 100 GeV Fermi LAT sensitivity does not exceed $10^{-13}$ erg cm$^{-2}$ s$^{-1}$. If we consider these curves together with the expected spectrum of the cloud (Fig. \ref{fig:spectrum}), we can estimate the maximum distance of detection as $1.3 \times 10^4$ AU.  
% It  could be seen that in case of greater distance the observed flux in $[1 - 100]$ GeV energy range would fall below all the sensitivity curves and Fermi LAT would not be able to detect such a cloud. 
This allows us to conclude that the number of molecular clouds should not exceed $\sim5 \times 10^2$ sources  closer than $r_{sens} \approx 13 \times 10^3 AU.$
This correspond to limit on the concentration:
\begin{equation}
n < 515 \times \frac{3}{4 \pi r_{sens}^3} \approx 6.3 \times 10^{-11} \text{AU}^{-3} \approx 5.6 \times 10^{5} \text{pc}^{-3}.
    \label{eq:soft_limit}
\end{equation}

Anyway, it is clear that it is rather impossible that all of the unassociated sources were dark molecular clouds. In the following, we use additional data to put another constraint on the population.

%%%%%%%%%%%%%%%%%%%%%%%%%%%%%%%%%%%%%%
%%%%%%%%%%% UV DATA AND MAIN LIMITATION %%%%%%%%%%%%%
%%%%%%%%%%%%%%%%%%%%%%%%%%%%%%%%%%%%%%

High column densities make cold molecular clouds opaque at wavelengths shorter than IR,  in the results of wide surveys at such wavelengths they would manifest themselves as dark circular shadows.  The size of the shadow would increase with increasing frequency, due to enhancement of the absorption, as the opacity increases.

We used  data from the all-sky UV-survey GALEX \citep{Morrissey}, which are optimal for our aims: it has sufficient angular resolution 6-\ang{;;8}, the source density is high, and, finally, it has  almost full sky coverage. 

    The GALEX telescope observed sky from 2003 to 2013 in two wavebands: near UV (NUV, 1750-2800 $\mathring{A}$) and far UV (FUV, 1350-1750 $\mathring{A}$). The corresponding angular resolution is  \ang{;;6.0} \ for the FUV channel  and \ang{;;8.0} \ for the NUV channel. GALEX peformed five imaging surveys, among which the All-Sky Imaging Survey (AIS) is the one with the largest area coverage (which reaches nearly 2/3 of the sky), with both FUV and NUV detectors exposed: it contains more than $10^7$ sources down to 20-21$^m$. 

GALEX observed sky in circular areas called tiles. The images of these areas are also called tiles. Each tile has 1.2 degrees angular diameter. AIS  comprises  of $5.7 \times 10^4$ observations of $2.9 \times 10^4$ fields. It means that each field was observed one or more times - in one or more "visit". Each visit  includes either  a single pointing or several adjacent pointings or "subvisits". The final catalogs of UV sources, such as \citep{galex}, used "coadd products",  i.e. co-added partial-exposure images or co-added "visits". In the present work we used (sub)visit-level products in order to avoid "blurring" of the cloud shadow by the parallax effect. 

% We first evaluated the size of the shadow. The column density corresponding to the optical depth to the absorption $\tau=1$ for different wavelength sets the  radii of opacity: $r_{FUV}= 0.4$~AU,  $r_{NUV}= 0.3$~AU. 
Molecular hydrogen is almost transparent at wavelength larger than 1100 nm (Lyman-Werner band), at larger wavelength absorption arises due to such hydrogen ions as $H_2^+$ and $H^-$ \citep{Heays2017}. Only the former one is relevant in our setting, it is produced in  processes of photoionization by cosmic rays and extreme UV photons (<800 nm), both primary and secondary \citep{Padovani2024}. Only $10^{-7}$($10^{-9}$) fraction of $H_2^+$ would be sufficient to make the cloud opaque at NUV(FUV) wavelength. Despite quick destruction of $H_2^+$ in interactions with $H_2$ molecules, needed concentrations could arise  because of   large photoionization rates due to secondary electrons \citep{Padovani2024}. Large concentrations of $H_2^+$ would be reached in the outer regions of the clouds already \citep{Varshalovich1993}, so we adopt $r_{UV}= 0.5$~AU for further estimates.\\

 We found the closest GALEX tile for each of the sources from our 4FGL-DR4 sample. 
 For some of the sources from the sample the closest tile did not contain the ellipse marking the 95\%-confidence position of the gamma-ray source. The total number of such sources is 107. They correspond to the sky regions not covered by AIS, so we did not consider them further. 
 
 In case the closest tile contained at least a part of an ellipse marking the 95\%-confidence position of a gamma-ray source, we downloaded all the available images of that tile. For each tile we downloaded from 1 to 8 images depending upon the number of visits and bands in which that region had been observed. Then we cropped a square region centered at the position of the corresponding source from all the downloaded images. The angular size of the region was twice as large as the major axis of the 95\%-confidence error ellipse. \hanna{The threshold of possible shadow detection is set by the density of the UV background, so that  on average the size of the detectable shadow could not be much lower than $\ang{;;50}$, otherwise the shadow would be lost in the fluctuations of the background. A shadow of size $r_{UV}$ would be noticeable up to the limiting distance of $<4\times10^{3}~$AU.} 

\begin{figure}%[h!]
\centering
\includegraphics[width=0.4\textwidth]{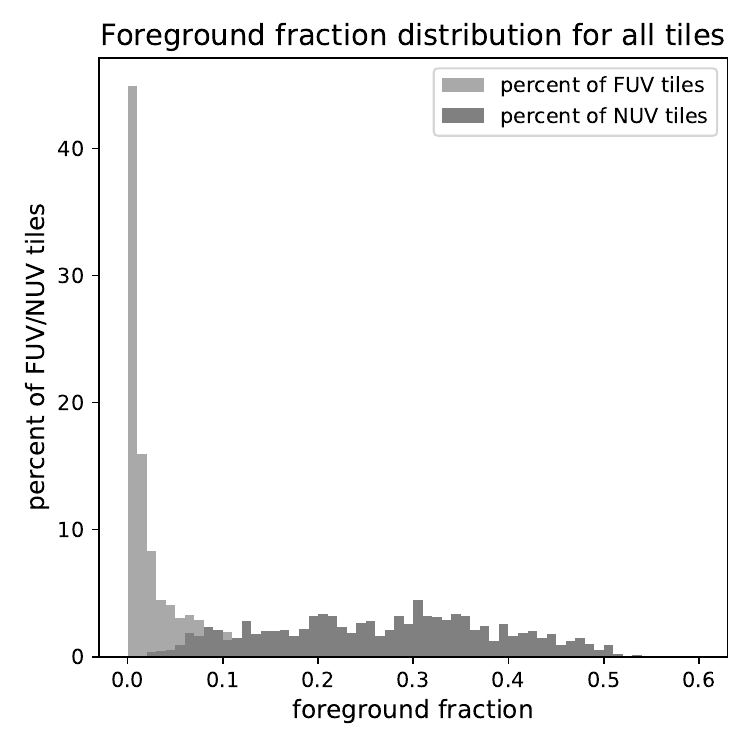}
\caption{The distribution of UV background by the contribution of the foreground (airglow + zodiacal light) for FUV and NUV tiles.}
\label{fig:foreground}
\end{figure}

The shadows we look for should not be completely black because of UV foreground originating from closer regions of the Solar System.  We estimated the expected signal from the shadow's area using the data on UV emission foreground by \citep{foreground}. The author attributes the foreground to two main UV sources: airglow from geocoronal oxygen in both FUV and NUV bands and zodiacal light in the NUV band only. He also presents \ang{;6} resolution maps of FUV and NUV background and detailed tables which contain the estimated airglow, zodiacal light, total FUV and NUV signal and a variety of other parameters for each pixel. \ang{;6} is just the mean angular size of the major axis of the error ellipses at 95\% confidence level. We calculated the contribution \hanna{$\alpha$} of the UV foreground to the total FUV and NUV fluxes \hanna{for each of the error ellipses from our sample}. The distribution of UV foreground contribution  to the total fluxes for FUV and NUV tiles is presented  in  Fig.\ref{fig:foreground}.

\begin{figure}
\centering

\begin{minipage}{.5\textwidth}
  \centering
  \includegraphics[width=.9\linewidth]{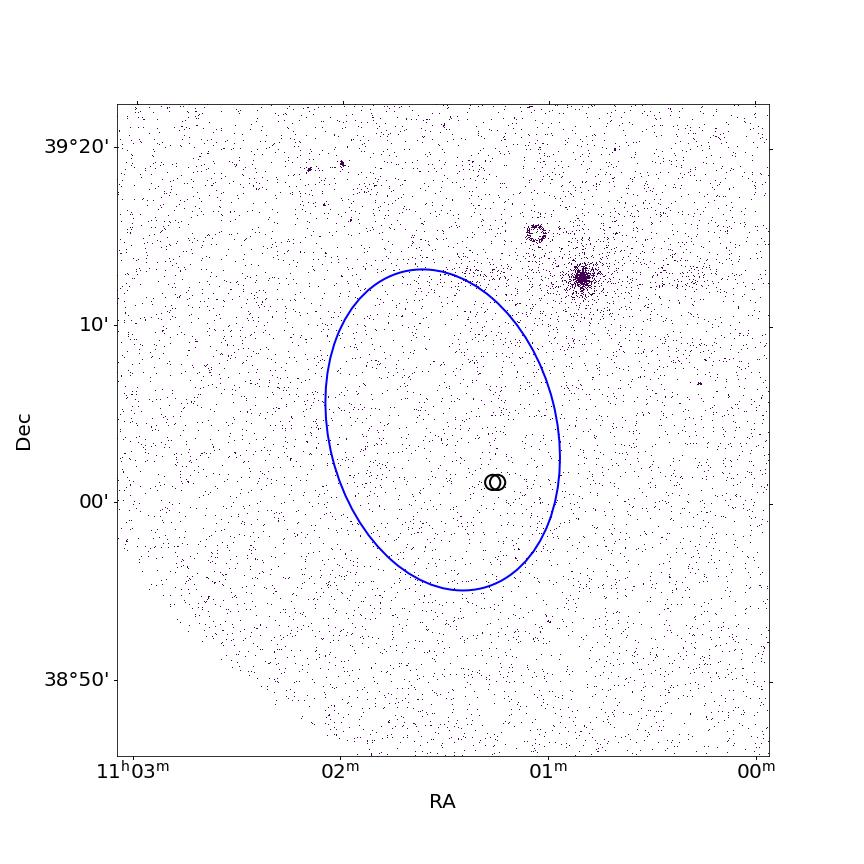}
  \caption{\hanna{A FUV image of a region cropped from tile AIS\_90\_0001\_sg13.\\
  The tile was observed on Feb 15, 2007 at 14:08:38. The exposure time is 106~s. \\
  The blue ellipse is the 95\%-confidence error ellipse of \\
  4FGL J1101.5+3904 gamma-ray source. \\
  The black circles indicate the position of a possible cloud shadow.}}
  \label{fig:fuv}
\end{minipage}%
\begin{minipage}{.5\textwidth}
  \centering
  \includegraphics[width=.9\linewidth]{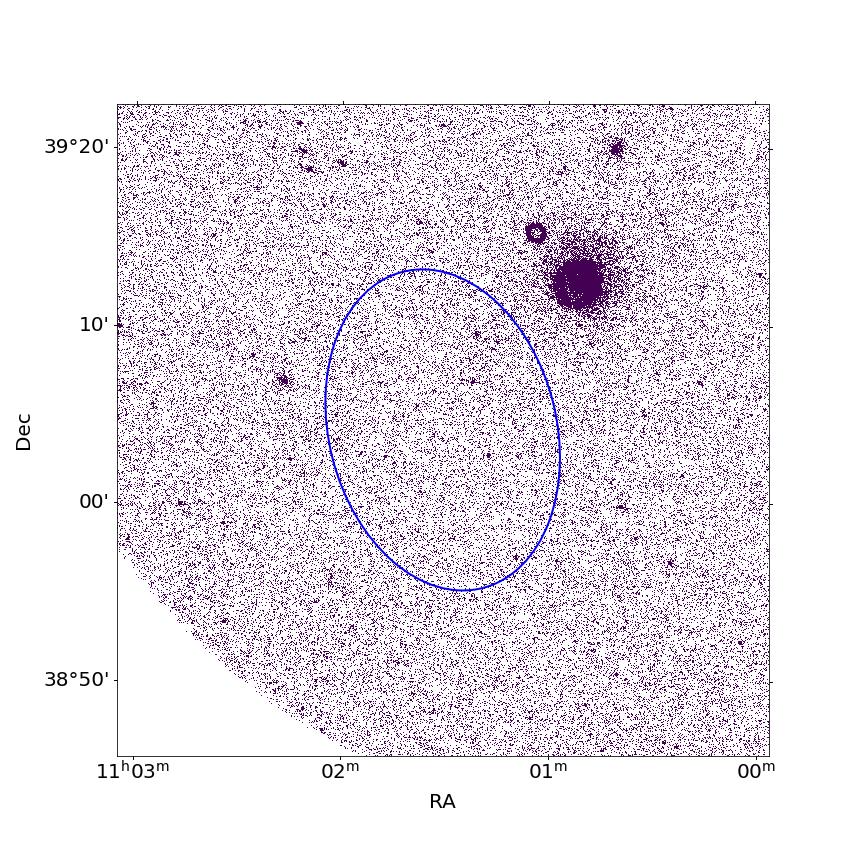}
  \caption{ \hanna{A NUV image of a region cropped from tile AIS\_90\_0001\_sg13.\\
  The tile was observed on Feb 15, 2007 at 14:08:38. The exposure time is 106~s. \\
  The blue ellipse is the 95\%-confidence error ellipse of \\
  4FGL J1101.5+3904 gamma-ray source. \\
  No dark features similar to that on the FUV image are present. }}
  \label{fig:nuv}
\end{minipage}%
\end{figure}

\begin{figure}
\centering

\begin{minipage}{.5\textwidth}
  \centering
  \includegraphics[width=.9\linewidth]{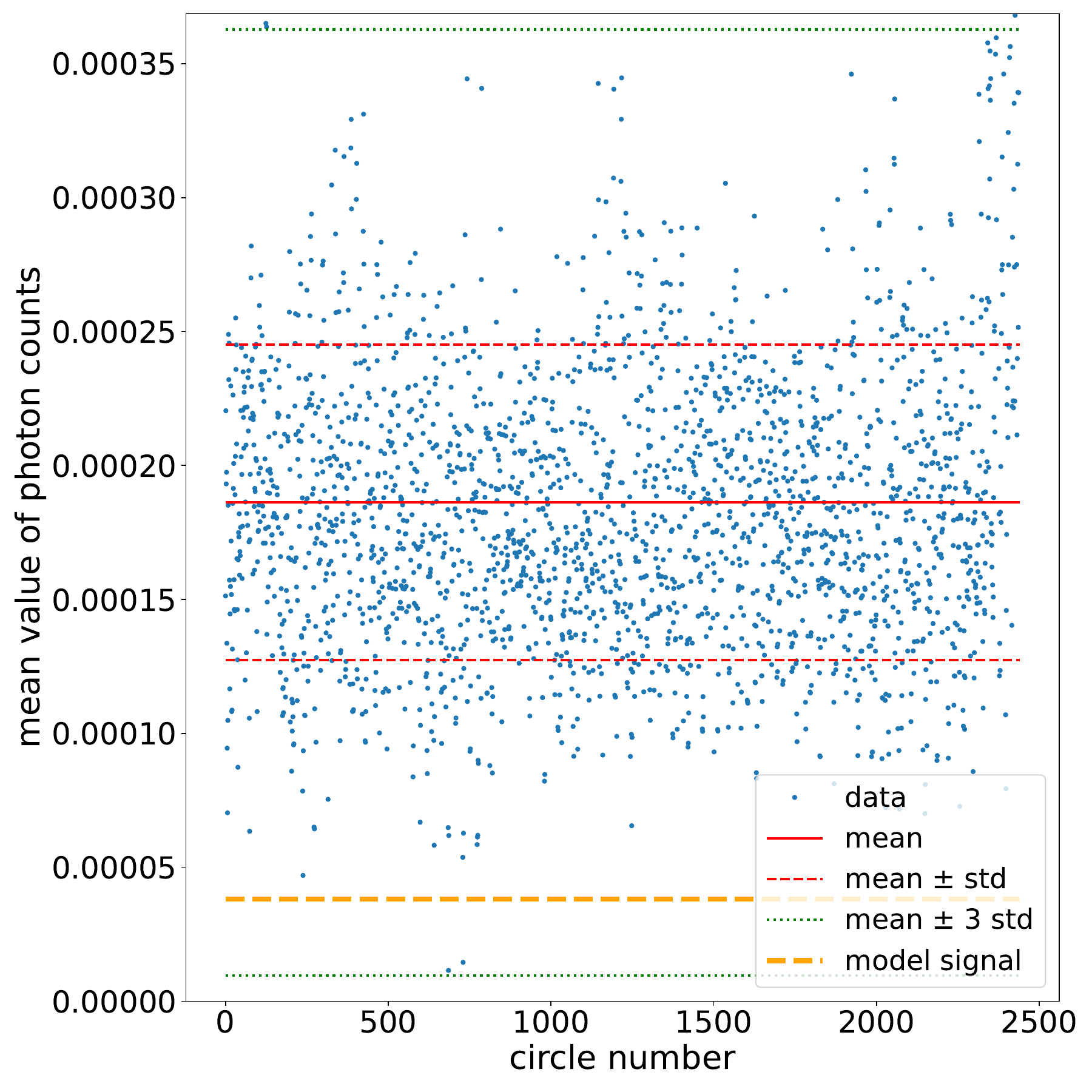}
  \caption{\hanna{The distribution of the signal from the circular regions for 
  FUV image \\in Fig.\ref{fig:fuv}; the red solid line indicates the mean of the distribution, 
  the red \\dashed lines indicate the mean $\pm ~\sigma_{dist}$ 
  and the green dashed-dotted lines \\indicate the mean $\pm ~3\sigma_{dist}$. 
  The thick orange line indicates the expected level of the signal from the shadow of a possible cloud. }}
  \label{fig:distfuv}
\end{minipage}%
\begin{minipage}{.5\textwidth}
  \centering
  \includegraphics[width=.9\linewidth]{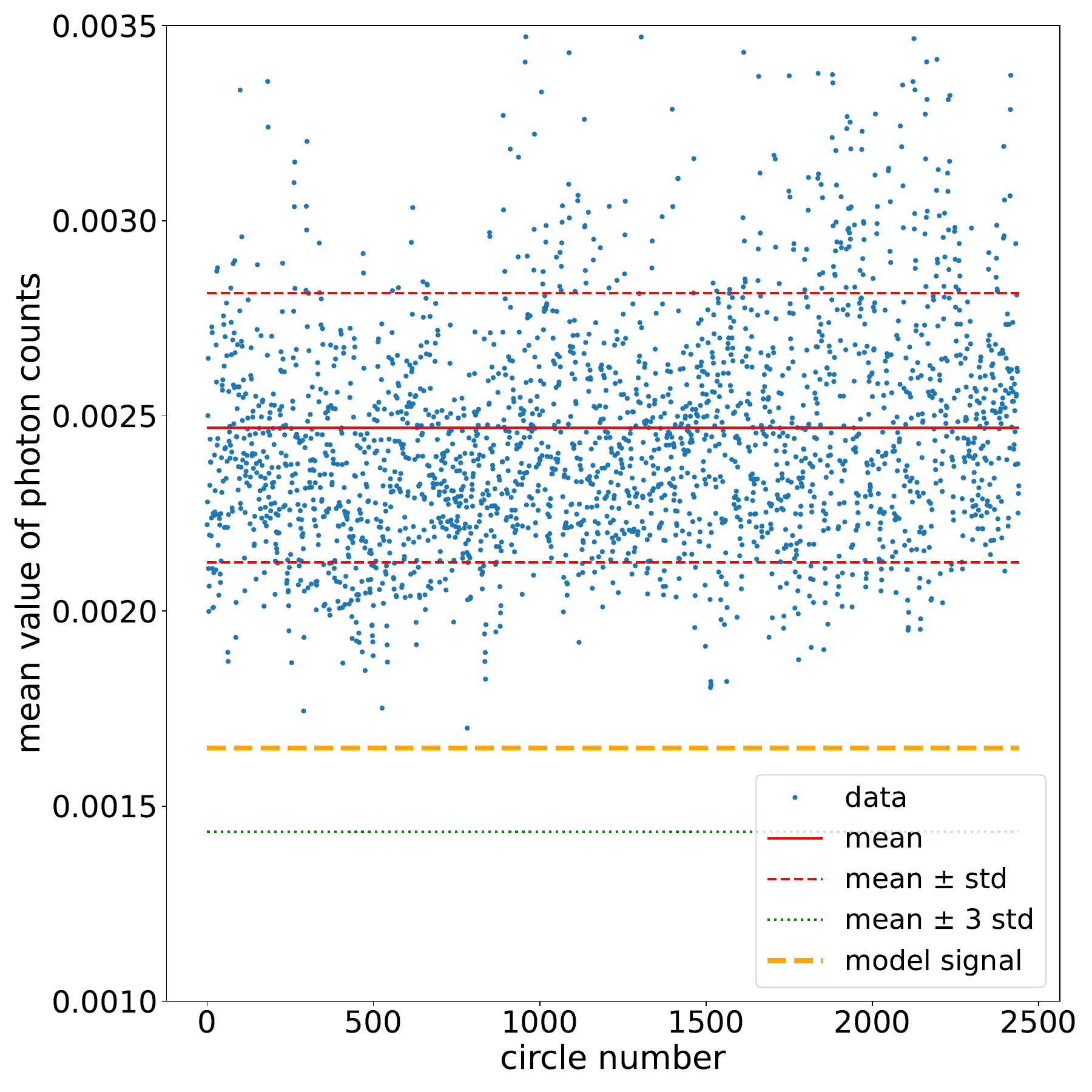}
  \caption{\hanna{Same as previous figure for NUV image in Fig.\ref{fig:nuv}.}}
  \label{fig:distnuv}
\end{minipage}%
\end{figure}

\begin{figure}
\centering
\begin{minipage}{.5\textwidth}
  \centering
  \includegraphics[width=.9\linewidth]{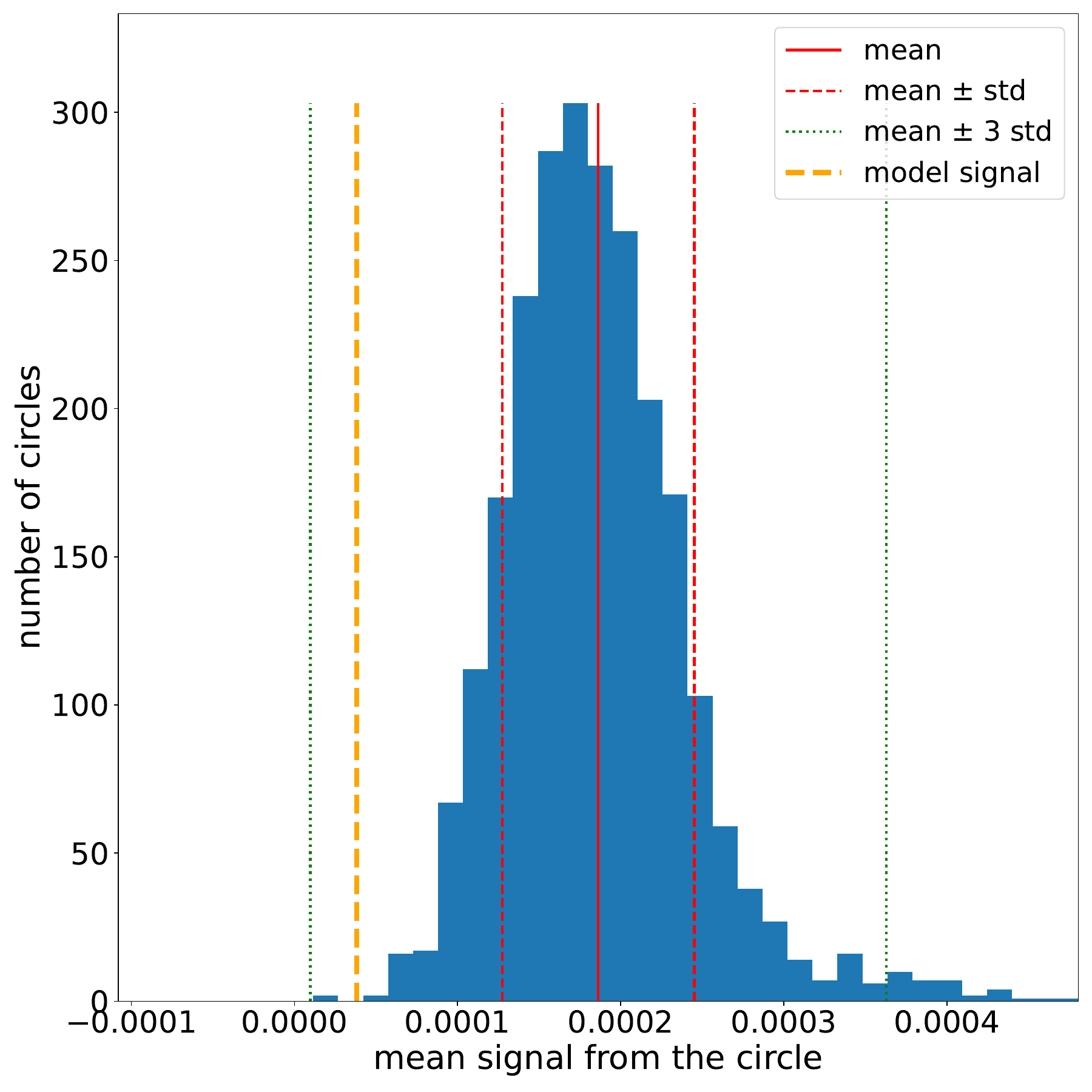}
  \caption{\hanna{The histogram of  the distribution of the signal  from the circular regions for FUV image in Fig.\ref{fig:fuv};  the designations are the same as in Fig. \ref{fig:distfuv}}}
  \label{fig:histfuv}
\end{minipage}%
\begin{minipage}{.5\textwidth}
  \centering
  \includegraphics[width=.9\linewidth]{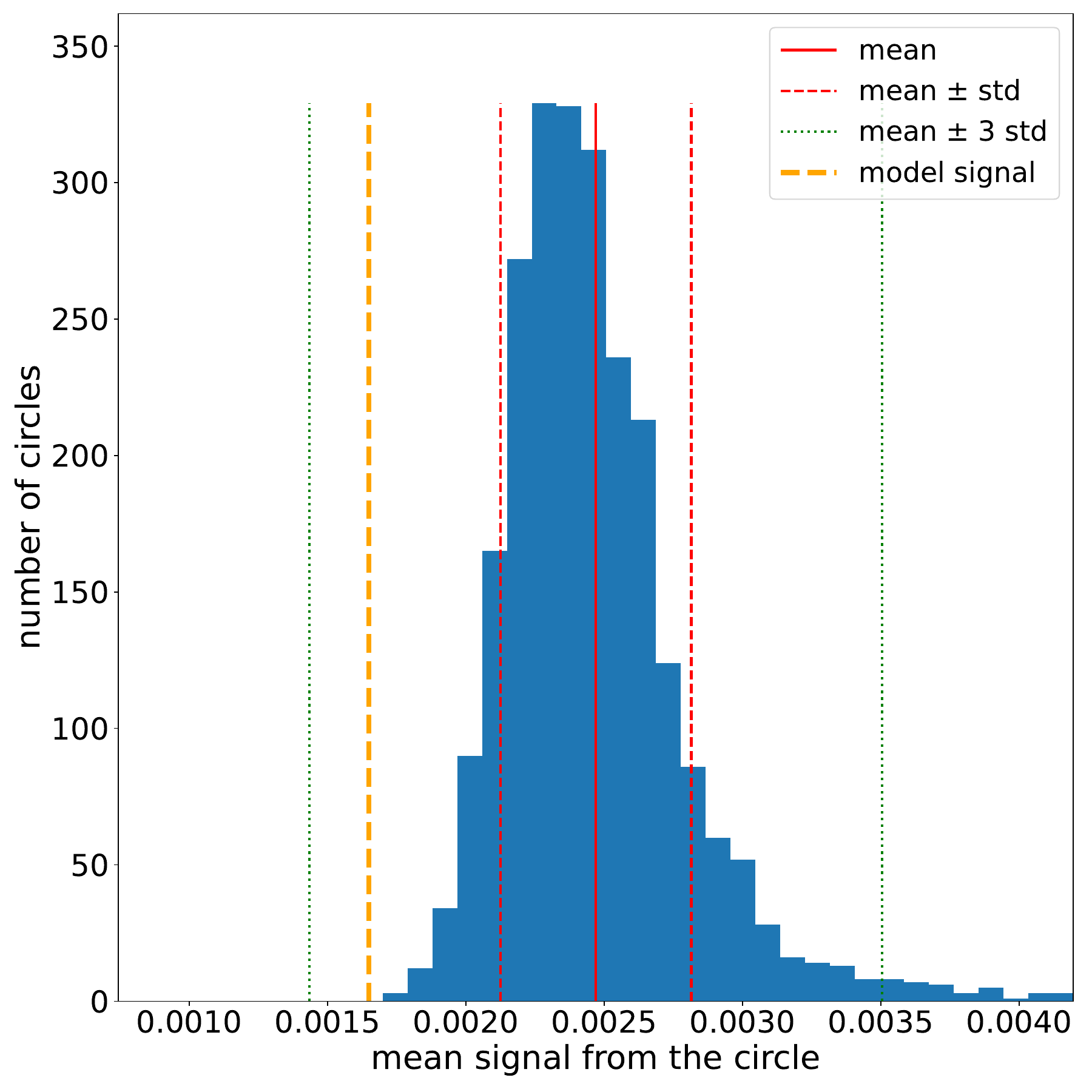}
  \caption{\hanna{Same as previous figure, but  for 
  NUV distribution.}}
  \label{fig:histnuv}
\end{minipage}%
\end{figure}

  \hanna{We automatically searched for dark circular shadows on sky regions corresponding to 95\%-confidence error ellipses of the selected 4FGL-DR4 sources. For each ellipse we cropped all the pixels brighter then $<f_{pix}> + ~3\sigma$, where $<f_{pix}>$ is the mean of all non-zero pixels in the ellipse and $\sigma$ is non-zero pixels' rms deviation. These pixels  likely correspond to the stars, while the others correspond to sky UV background.  
  
  We assumed the size of sought-for circular regions to be $r_{shad} \approx $ \ang{;;25.8}- that is just the expected size of the model cloud at a distance of $4\times10^{3}~$AU. The FWHM of GALEX PSF is \ang{;;4.9} and \ang{;;4.2} for NUV and FUV bands respectively, this may lead to additional blurring of the shadow. We divided the shadow into two regions. The central region of radius $(r_{shad} - r_{PSF})$, where $r_{PSF} = $ \ang{;;4.9} for NUV data and $r_{PSF} = $ \ang{;;4.2} for FUV data, is not affected by PSF smearing, thus the expected flux density from this region is of the mean foreground level: $<F_{cen}>~ \approx~ \alpha <F_{UV}>$. The flux density from the second region, which is a ring of $r_{PSF}$ width, was roughly estimated as the mean of the foreground level and the mean level of background flux density typical for the corresponding tile: $<F_{ring}> ~\approx~ <F_{UV}> \cdot \frac{1 + \alpha}{2}$. The mean flux density of the whole possible shadow was evaluated as a sum of these two contributions: 
  \begin{equation}
      F_{shad} ~\approx~ \alpha <F_{UV}> \frac{(r_{shad} - r_{PSF})^2}{r_{shad}^2} +   \frac{1 + \alpha}{2} <F_{UV}> \frac{r_{PSF} (2r - r_{PSF})}{r^2} .
  \end{equation}
  
  Then we looked whether there were any circular regions with both low enough mean flux density ($\leq F_{shad}$) and dark enough  central region ($ \leq (\alpha~ \cdot <F_{UV}>)$). If any particular region which satisfy these criteria had been found, that region would have been selected for a further check - we looked for the same deviation on the other images of that tile. The number of such "suspicious" regions turned out to be not large: less then 3\% of all tiles showed such dark features. All of the tiles with such features correspond to the FUV band, where the density of the background is noticeably lower. None of such "suspicious" regions showed dark counterparts on the NUV tiles observed on the same date. This leads to the conclusion that these dark circles are just random fluctuations of the UV background.  

  We have also built the distributions of the mean signal from all the circular regions considered for each ellipse and the histograms of  these distributions. The vast majority of the points of these distributions lie within 3 standard deviations $\sigma_{dist}$ from the mean, and most of the outliers with lower signal correspond to the circular regions located at the edge of GALEX tile (this is the case when the tile contains only a part of 4FGL-DR4 95\%-confidence error ellipse of the corresponding source and the outliers arise because of the lack of GALEX data). Rare points with deviations larger then $3 \sigma_{dist}$ which do not correspond to the edge of GALEX tile also do not stand out very significantly from the others. In general the distributions and histograms  confirm the assumption that the outliers are just more prominent fluctuations and we can conclude  that there are no dark molecular clouds closer than $4\times10^{3}~$AU.
 
 An example of data analysis  of  one of the considered regions is presented in Fig.\ref{fig:fuv}-\ref{fig:histnuv}.  Fig. \ref{fig:fuv} shows a FUV image of a region cropped from tile AIS\_90\_0001\_sg13. The tile was observed on Feb 15, 2007 at 14:08:38. The exposure time was 106~s.  Fig. \ref{fig:nuv} shows a NUV image of the the same region observed simultaneously with the first one. The blue ellipse is the 95\%-confidence error ellipse of 4FGL J1101.5+3904 gamma-ray source. The black circles  in the FUV image (Fig.\ref{fig:fuv}) indicate the position of a possible cloud shadow. The absence of such features in the NUV image(Fig.\ref{fig:nuv}) lead to the proposal that it  was a fluctuation. Fig.\ref{fig:distfuv} and Fig.\ref{fig:distnuv} which present the distributions of the signal from the circular regions for NUV and FUV tile's regions respectively, confirm the assumption that the dark region is a fluctuation; the red solid line indicates the mean of the distribution, the red dashed lines indicate the mean $\pm ~\sigma_{dist}$ and the green dotted lines indicate the mean $\pm ~3\sigma_{dist}$. The thick orange line indicates the mean signal expected from the shadow of a possible cloud. Fig.\ref{fig:histfuv} and Fig.\ref{fig:histnuv} show the histograms for these distributions.}

If we assume that the clouds in the inner regions of the Solar System are distributed uniformly enough and that independent measurements of the number of clouds in a specific volume follow Poisson distribution, the fact that we got 
no clouds' shadows in the AIS Survey means that with 95\% confidence the upper limit on the number of clouds found in the volume defined by solid angle $\Omega_{GALEX} \approx 2.6 \times 10^4 ~ \text{deg}^2$ of GALEX sky coverage and our limiting distance \hanna{$r \approx 4 \times 10^3$ AU } is $N_{max} \approx 3.7$ clouds. This allows to constrain  the concentration of dark molecular clouds in the inner regions of the Solar System:

\begin{equation}
\hanna{n < N_{max} \times \frac{3}{\Omega_{GALEX} r^3} \approx 2.2 \times 10^{-11} \text{AU}^{-3}\approx 2.0 \times 10^{5}} \text{pc}^{-3}
    \label{eq:limit_AU}
\end{equation}

%%%%%%%%%%%%%%%%%%%%%%%%%%%%%%%%%%%%%%
%%%%%%%%%%% PARALLAX AND PROPER MOTION %%%%%%%%%%%%%
%%%%%%%%%%%%%%%%%%%%%%%%%%%%%%%%%%%%%%

\section{Discussion}\label{sec:discussion}
Positional shift due to the parallax should also be considered in the analysis. If it exceeds the size of Fermi-LAT error ellipse the object would not be detected as a  point-like gamma-ray source. Sizes of error ellipses in our sample vary from \ang{;1.4} to \ang{;31} with the mean value of $(\ang{;6.0} \pm \ang{;3.6})$. It means that Fermi-LAT can not detect a cloud as a point-source object if it is situated closer then in average $\approx 6 \times 10^2$ AU.

Parallax shift can also be the cause of blurring effect: as  its value exceeds the size of the cloud, stacking of different epochs would  disrupt the images of the shadows on GALEX tiles. In order to avoid such problems, we used single-epoch observations. That is why we used only visit- and subvisit-level data from GALEX data products.  

Another possible source of confusion could be the proper motion of the cloud. GALEX observatory observed sky from 2003 to 2013, while Fermi LAT operated since 2008, so the typical timescale is 10 years. Angular displacement in 10 years should not be greater then the size of Fermi-LAT error ellipse, otherwise both the cloud would not be identified as a point source and its shadow would not get inside the error ellipse. As stated above we are interested only in gravitationally bound clouds that perform Keplerian motion around the Sun, $v_{cl}=\sqrt{\frac{GM_{\odot}}{d}} \approx 1 \left(\frac{d}{1000~\mathrm{AU}}\right)^{-1/2}~\mathrm{km~s^{-1}}, \mu_{cl}=\frac{1}{d} \sqrt{\frac{GM_{\odot}}{d}} \approx 40'' \left(\frac{d}{1000~\mathrm{AU}} \right)^{-3/2}~\mathrm{yr}^{-1}$, so for $d>1000$~AU angular displacement in 10 years  would be considerably smaller than the size of the error ellipse, which shows that the point-like approximation could be used.
It would not be the case for the clouds belonging to the halo population with $v\sim300~\mathrm{km~s^{-1}}$, that is why we limited our analysis to the bound subpopulation.
Parallax and proper motion constraints slightly reduce the effective volume of our approach, $\Omega_{GALEX} r^3 \rightarrow \Omega_{GALEX} (r_{max}^3-r_{min}^3)$, where $r_{max}=3$~kpc and $r_{min}=1$~kpc.

%%%%%%%%%%%%%%%%%%%%%%%%%%%%%%%%%%%%%%
%%%%%%%%%%% UV FOREGROUND %%%%%%%%%%%%%
%%%%%%%%%%%%%%%%%%%%%%%%%%%%%%%%%%%%%%

\section{Results \& Conclusions} \label{sec:results}
We put the limits on the concentration of small dark clouds in the solar system in two separate ways. 

In the first method we estimated the gamma-ray luminosity of the cloud and thus the volume in which the detection of such cloud by Fermi-LAT is possible. We evaluated the luminosity of the model cloud in the energy range from 1 to 100 GeV as $6.8\times10^{25}~\mathrm{phot~s^{-1}}$ and this together with Fermi LAT sensitivity curves allowed us to put a constraint on clouds' concentration supposing that all of unidentified sources are dark clouds: $n < 6.3 \times 10^{-11} {\mathrm{AU}^{-3}} $.

In the second method we looked for the dark shadows left by the clouds on sky UV background. The angular size of the detectable shadows should be greater than the typical angular size of the fluctuations in the GALEX background data. Our analysis with automated search of such dark regions on GALEX data allowed us to conclude that there are no clouds closer than \hanna{$4 \times 10^3$ AU}. Taking into account the constraints set by parallax shift and proper motion of the cloud, we got the upper limit on the concentration of the clouds in the inner regions of the Solar System :\hanna{ $n < 2.2 \times 10^{-11} {\text{AU}^{-3}} \approx 2.0 \times 10^5 ~{\text{pc}^{-3}}$} .

We can see that the second method provides us with a stricter constraint on the concentration of dark clouds in the inner regions of the Solar System. 

If we suppose that the clouds are distributed uniformly inside the Solar Hill surface, we can estimate the upper limit on the total number of such objects bound to the Sun. Assuming that for the Sun the Hill surface is an ellipsoid with principal semi-axes of (1.42, 0.92, 0.74) pc, in  cylindrical Galactic  coordinates \citep{Hill_surface}, we achieve:
\begin{equation}
\hanna{N_{clouds} <  7.9 \times 10^{5}}
    \label{eq:N_full}
\end{equation}

The upper limit on the total mass of these objects in case of uniform distribution of the clouds:
\begin{equation}
\hanna{M_{clouds} <  30 M_{\odot}}
    \label{eq:M_full}
\end{equation}

Due to several effects, we would expect a more concentrated distribution of the clouds. First, some clouds in the outer regions of the system were lost due to the interactions during stellar flybys in the last 5 Gyr. Second, the collapse of the gas cloud into a more compact star leads to the adiabatic contraction of gravitationally bound population of the dark clouds. If the latter contribution to the total mass was subdominant, then a rather spiky profile would have been formed, $n(r)\sim r^{-3/2}$ \citep{Capela2013}. In this case the total number of clouds grows  with a radius much slower, $N\sim r^{3/2}$, not $N\sim r^{3}$  and total number of bound clouds and their total mass would be constrained much stronger: 
\hanna{
\begin{alignat}{2}
&N_{clouds} <  1.7 \times 10^3, \\%10^{4},\\
&M_{clouds} <  0.05~M_{\odot}.%0.3~M_{\odot}.
    \label{eq:NM_full_AC}
\end{alignat}   
}

%  \begin{eqnarray}
% N_{clouds} <  10^{4},\\
% M_{clouds} <  0.3~M_{\odot}.
%     \label{eq:NM_full_AC}
% \end{eqnarray}   

There is no general solution when the masses of components (gas/dark clouds) are close, one would expect some degree of contraction, but the resulting profile would be shallower than $n(r)\sim r^{-3/2}$. The method of UV-shadows is more constraining in case of contracted profiles, as we have there strong exclusion limits for the inner region where now comparatively larger fraction of the total population resides.

In conclusion we point out that future all-sky UV or X-ray missions with better angular resolution, longer exposure and better sky coverage are required to improve the result and to set  stricter constraints on dark molecular clouds' concentration in the Solar System.

\section*{Acknowledgements}
We would like to thank Alexander Popkov for useful comments.
The authors would also like to thank  anonymous referees whose comments and suggestions helped to significantly improve the quality of the paper. 
The work of the authors was supported by the Ministry of Science and Higher Education of Russian Federation
under the contract 075-15-2024-541 in the framework of the Large Scientific Projects program within the national project "Science". This research has made use of NASA’s Astrophysics Data System.

%% Bibliography
%% Author year style
\bibliographystyle{jasr-model5-names}
\biboptions{authoryear}
\bibliography{refs}

\end{document}